\documentclass[english,prd,eqsecnum,nofootinbib,floatfix]{revtex4}
\usepackage[T1]{fontenc}
\usepackage[latin9]{inputenc}
\usepackage{amsmath}
\usepackage{amssymb}
\usepackage{esint}

\makeatletter
\@ifundefined{textcolor}{}
{%
 \definecolor{BLACK}{gray}{0}
 \definecolor{WHITE}{gray}{1}
 \definecolor{RED}{rgb}{1,0,0}
 \definecolor{GREEN}{rgb}{0,1,0}
 \definecolor{BLUE}{rgb}{0,0,1}
 \definecolor{CYAN}{cmyk}{1,0,0,0}
 \definecolor{MAGENTA}{cmyk}{0,1,0,0}
 \definecolor{YELLOW}{cmyk}{0,0,1,0}
 }

\usepackage{amsfonts}
\makeatother

\usepackage{babel}

\makeatother

\usepackage{babel}

\makeatother

\usepackage{babel}

\makeatother

\usepackage{babel}

\makeatother

\usepackage{babel}

\makeatother

\usepackage{babel}

\begin{document}
KCL-PH-TH/2010-20  

\vspace{10mm}

\title{Stringy Space-Time Foam, Finsler-like Metrics and Dark Matter Relics}

\author{Nick E. Mavromatos, Sarben Sarkar and Ariadne Vergou}

\affiliation{King's College London, Department of Physics, Strand WC2R 2LS, London,
U.K.}
\begin{abstract}
We discuss modifications of the thermal Dark Matter (DM) relic abundances
in stringy cosmologies with D-particle space-time foamy backgrounds.
As a result of back-reaction of massive DM on the background space-time,
owing to its interaction with D-particle defects in the foam, quantum
fluctuations are induced in the space-time metric. We demonstrate
that these lead to the presence of extra source terms in the Boltzmann
equation used to determine the thermal dark matter relic abundances.
The source terms are determined by the specific form of the induced
metric deformations; the latter depend on the momentum transfer of
the DM particle during its interactions with the D-particle defects
and so are akin to Finsler metrics. In the case of low string scales
arising from large extra dimensions our results may have phenomenological
implications for the search of viable supersymmetric models.
\end{abstract}
\maketitle

\section{Introduction}

The nature of the Dark sector of our Universe constitutes one of the
major unresolved puzzles of modern physics. Indeed, according to observations
over the past twelve years, 96 \% of our Universe energy budget consists
of unknown entities: 23\% is Dark Matter (DM) and 73 \% is dark energy
(DE), a mysterious form of ground state energy. DE is believed to
be responsible for the current-era acceleration of the Universe. These
numbers have been obtained by best-fit analyses of a plethora of astrophysical
data to the so-called Standard Cosmological Model ($\Lambda$CDM),
which is a Friedman-Robertson-Walker (FRW) cosmology, involving cold
DM, as the dominant DM species, and a positive cosmological constant
$\Lambda>0$; the data range from direct observations of the Universe
acceleration, using type-Ia supernovae~\cite{snIa}, to cosmic microwave
background~\cite{cmb}, baryon oscillation~\cite{bao} and weak
lensing data~\cite{lensing}. It should be stressed that the afore-mentioned
energy budget depends crucially on the theoretical model for the Universe
considered.

An interesting, and not commonly discussed, class of Cosmological
models that may lead to modifications of the Dark sector, involves
space-time with a {}``foamy'' structure at microscopic (Planckian
or string) scales~\cite{wheeler}, due to quantum gravitational interactions.
In the past, for a variety of reasons, such models (differing in the
details of the constructs of space-time foam) have been considered
by many authors. They exhibit a profusion of features that can be
falsified experimentally and their predictions range from light-cone
fluctuations caused by stochastic metric fluctuations~\cite{ford},
to macroscopic Lorentz symmetry violations~\cite{lorentz}. In this
note we would like to present a first study towards the contribution
of such space-time metric stochastic fluctuations on the Dark sector
and in particular on the DM sector of the Universe. We shall focus
on a particular class of stochastic space-time foam models, inspired
by certain types of string theory. These involve localized space-time
defects (D0-branes or D-particles)~\cite{dfoam} which are either
allowed background configurations~\cite{dfoam,emnnewuncert} or arise
effectively from suitable compactifications of higher dimensional
branes (e.g. D3-branes wrapped up in appropriate three cycles in the
context of type IIB strings~\cite{li}) . Observers on the brane
detect a foam-like structure due to the crossing of the brane by D-particles.
In this higher-dimensional geometry, only gravitational fields are
allowed to propagate in the bulk; all other particle excitations,
including DM candidates, are assumed to be described by open strings
with their ends attached to the brane world. The brane is assumed
to have three large spatial dimensions, and -- depending on the model
of string theory considered -- it may have a number of compactified
extra dimensions.

Dynamical D-particles should\emph{ not} be viewed as material excitations
of the vacuum but rather as \emph{vacuum structures}. This contrasts
with attempts to represent such D-particles as ordinary superheavy
DM excitations from the vacuum~\cite{d0dm}, owing to the completely
localized nature of the D-particles in the extra dimensions. In our
construction, closest in spirit to weak coupling string theory, they
are just \emph{vacuum defects}. In fact it can be shown that the gravitational
interactions among such (BPS) D-particles are cancelled by appropriate
gauged repulsive forces induced on them from other branes in our supersymmetric
models of D-foam~\cite{dfoam}. Hence such a collection of D-particles
in the bulk does not affect the Hubble expansion on the D3-brane worlds,
and so their concentration cannot be restricted by considerations
on overclosure of the Universe within our models. Important constraints
on the density of defects in D-foam models can still be imposed by
astrophysical experiments on the arrival times of high energy cosmic
photons~\cite{arrival}. According to these string-foam models, there
should be an effective \emph{refractive} index in vacuo, such that
higher energy photons would be delayed more, since they would cause
stronger disturbance on the background space-time~\cite{dfoam,emnnewuncert,li}.
The interaction of material open strings (such as photons) with the
D-particles leads to an induced distortion of space-time described
by a metric, which depends on both the coordinate and momentum transfer
of the photon during its scattering with the defect and so has similarities
to a Finsler metric~\cite{finsler}. This is a topologically non-trivial
process, involving the creation of a non-local intermediate string
state, oscillating from zero length to a maximum one, according to
a time-space stringy uncertainty~\cite{sussk1,emnnewuncert}. This
causes a time delay for the photon emerging after capture by the defect,
proportional to the incident energy of the photon. .

The purpose here is to analyze, in the same spirit, the modification
of the estimate of the DM budget of the Universe as compared to the
$\Lambda$CDM model due to the quantum fluctuations of the D-particles.
In fact, as we shall argue, the propagation of massive DM particles
on such space-times, induces a back-reaction, which in turn has consequences
for the amount of thermal DM relics of these particles; this, in turn,
impacts on astro-particle tests of particle physics models incorporating
supersymmetry, that provide currently one of the leading DM candidate
species, the neutralino. Its thermal abundance, calculated within
the simplest supersymmetry models (minimal supersymmetric model embedded
in minimal supergravity~\cite{msugra}), is heavily restricted by
cosmic microwave background data; hence the available parameter space
for these simplest supersymmetric models may vanish in the near future,
on incorporating also data from collider experiments such as the LHC
at CERN~\cite{neutralino}. These constraints depend strongly on
details of theoretical models. In the presence, for instance, of extended
supergravity models with time-dependent dilaton-$\phi$ sources~\cite{elmn},
the calculated amount of thermal neutralino relic abundance can be
smaller than the one calculated within the $\Lambda$CDM-minimal supergravity
cosmology. Such dilaton models allow more scope for supersymmetry,
which can thus survive the otherwise stringent tests at the LHC~\cite{dutta}.
In our work we will find that the effects of the D-foam on the thermal
relic abundances oppose those from the dilaton models. These effects
may become relevant for models allowing low string mass scales.

For clarity we will first outline the basic reason behind such modifications
in the DM thermal abundances. Non-equilibrium cosmology models are
associated with space-time distortions, due to either the presence
of time dependent dilaton sources (cf supercritical (SSC) dilaton
quintessence string cosmologies~\cite{elmn}), or the induced back
reaction of the DM particles onto the space-time itself. Boltzmann
equations are used to determines the thermal DM species cosmic abundances.
The effect of space-time distortions can be subsumed as extra contributions
to a source $\Gamma(t)$ for particle production (at cosmic Robertson-Walker
time $t$) on the right-hand side of the appropriate Boltzmann equation.
In general (for a Universe with three spatial dimensions (i.e. a 3-brane),
with no loss of particles to the bulk ), with $n(t)$ the density
of the DM species, the apposite Boltzmann equation reads : \begin{equation}
\frac{d\, n}{dt}+3H\, n=\Gamma(t)\, n+\mathcal{C}[n]\label{boltzmann}\end{equation}
 where $H=\frac{\frac{da}{dt}}{a}\equiv\frac{\dot{a}}{a}$ is the
Hubble ratio, $a$ being the scale factor in the Robertson-Walker
metric, and $\mathcal{C}[n]$ is the standard collision term describing
the deviations of the species population from thermal equilibrium.
In standard cosmologies, which we assume here, this has the form~\cite{kolb}
\begin{equation}
\mathcal{C}[n]=-\langle{\tilde{\sigma}}\, v\rangle\left(n^{2}-n_{{\rm eq}}^{2}\right)\end{equation}
 with $n_{{\rm eq}}$ a thermal equilibrium density of species, ${\tilde{\sigma}}$
the total cross section evaluated in the background metric, and $v$
the Mœller velocity.

We will provide an explicit expression for the D-foam induced source
$\Gamma(t)$. The interaction of DM particles, represented as open
string excitations on the D3 brane world, with D-particles, (interpreted
as space-time defects in the ground-state ), results in metric distortions.
The latter depend on the momentum transfer of the matter particles
and so are of Finsler type~\cite{finsler}. It was observed that$\,$\cite{gravanis},
for the universe to expand by a scale factor $a(t)$, in a way consistent
with the world-sheet logarithmic conformal invariance of the D-particle
recoil process during capture and scattering~\cite{kogan}, there
is an induced space-time distortion which has the form: \begin{equation}
ds^{2}=-dt^{2}+v_{i}a^{2}(t_{0})dtdx^{i}+a^{2}(t)dx^{i}dx^{i}\delta_{ij}~,\quad t>t_{0}~.\label{background}\end{equation}
 Here the underlying space-time is given by a \emph{spatially-flat}
Robertson-Walker metric, which is the space observed by a low-energy
observer on the brane world. In (\ref{background}), $g_{s}$ is the
string coupling, $M_{s}$ is the string state, and $M_{s}/g_{s}$
is the mass scale of the quantum gravitational foam fluctuations in
the model. The metric (\ref{background}) has precisely the form corresponding
to a boosted frame (the D-particle's rest frame), with the boost occurring
suddenly at time $t=t_{0}$, the time of the capture of the string
state by the D-particle. On account of momentum conservation during
the scattering~\cite{kogan}, the D-particle recoil velocity $v_{i}$
is related to the momentum transfer $\Delta p_{i}$ of the string
state, which in turn can be parametrized as a fraction of the incident
momentum\cite{Liouville} \begin{equation}
v_{i}=g_{s}\frac{\Delta p_{i}}{M_{s}}\equiv r_{i}p_{i}\equiv g_{s}\frac{\xi_{i}p_{i}}{M_{s}}~,\qquad{\rm no~sum~over~i=1,2,3}~,\, r_{i}<1~,\label{defrecvel}\end{equation}
 where $p_{i}$ is a co-variant co-moving momentum of the DM particle
in the (distorted) Robertson-Walker background. In our foam model
we assume that the population of D-particles, is approximately uniform
and relatively dense over a given epoch of the Universe. In such a
case one may average (\ref{background}) using appropriate distribution
functions. Denoting such an average by $\ll\dots\gg$, and assuming
stochastic Gaussian distributions, \begin{equation}
\ll r_{i}\gg=0~,\qquad\ll r_{i}r_{j}\gg=\sigma_{i}^{2}\delta_{ij}~,\qquad{\rm no~sum~over~i=1,2,3}~.\label{average}\end{equation}
 The variances $\sigma_{i}^{2}$ need not be independent of $i$;
however to keep in line with the observed isotropy of the Universe
at large scales, such potential anisotropies (due to an anisotropy
of the population of D-particles in the bulk~\cite{dfoam,emnnewuncert})
would have to be small.

Upon considering such foam populations, where the time scale between
D-particle captures and emissions is much shorter than the inverse
mass of the DM, a coarse-grained description is applicable; the propagation
of a DM distribution over such a background is described on average
by metrics $\ll ds^{2}\gg$ in which to a good approximation $t_{0}\sim t$.
Consequently we write the distorted metric in the following form:
\begin{equation}
g_{\mu\nu}=\left(\begin{array}{cccc}
-1 & a^{2}(t)r_{1}p_{1} & a(t)^{2}r_{2}p_{2} & a^{2}(t)r_{3}p_{3}\\
a^{2}(t)r_{1}p_{1} & a^{2}(t) & 0 & 0\\
a^{2}(t)r_{2}p_{2} & 0 & a^{2}(t) & 0\\
a^{2}(t)r_{3}p_{3} & 0 & 0 & a^{2}(t)\end{array}\right)~.\label{metric2}\end{equation}
 The averaging over foam populations is performed using (\ref{average});
hence, when considering the evolution of DM densities in such backgrounds,
terms with an odd number of $r_{i}$'s will be ignored. This should
be understood in our subsequent discussion. The momentum dependence
of the metric (\ref{metric2}) implies a modification of the form
of the geodesic equation for $\frac{d^{2}x^{\mu}}{d\tau^{2}}$ together
with the modification of the Christoffel symbols. Keeping terms up
to order $r^{2}$ and dropping everywhere cross terms of the form
$r_{i}r_{j}$ for $i\neq j$ (these terms will in any case vanish
when one takes the average), it is straightforward to show that the
pertinent geodesic equation for $\mu=i\left(=1,2,3\right)$ reads:
\begin{eqnarray}
 &  & \frac{d^{2}x^{i}}{d\tau^{2}}=-\frac{2}{m^{2}}Hp^{i}p^{0}-\frac{2}{m^{2}}a^{2}(t)Hr_{i}p^{i}\left(p^{0}\right)^{2}+\frac{8}{m^{2}}a^{6}(t)Hr_{i}^{2}\left(p^{i}\right)^{3}p^{0}\nonumber \\
 &  & +\frac{2}{m^{2}}a^{4}(t)Hr_{i}p^{i}\sum_{j}\left(p^{j}\right)^{2}+\frac{4}{m^{2}}a^{4}(t)Hr_{i}^{2}p^{i}\left(p^{0}\right)^{3}-\frac{4}{m^{2}}a^{6}(t)Hr_{i}^{2}p^{i}p^{0}\sum_{j}\left(p^{j}\right)^{2}\end{eqnarray}
 where the Einstein summation convention is \emph{not} applied.

To discuss DM relics we need to solve the relevant Boltzmann equation
that describes the evolution of the phase-space density function of
the DM species in $t$. The distribution function $f$ of a particle
species is specified by phase space variables of the system; it is
thus natural to define a \emph{local} space-time momentum in an expanding
universe \cite{bernstein}: \begin{equation}
\overline{p}^{i}\equiv a(t)p^{i}~,\qquad i=1,2,3~.\label{bernstein}\end{equation}
 In terms of these scaled momenta, the energy-momentum dispersion
relation for a DM particle of mass $m$ in our (spatially flat Robertson-Walker
space-time background), assumes an effectively {}``Minkowski-space-time''
form, $p^{\mu}p^{\nu}g_{\mu\nu}=-E^{2}+a^{2}(t)p^{i}p^{j}\delta_{ij}=-E^{2}+\overline{p}^{i}\overline{p}^{j}\delta_{ij}=-m^{2}$.
Thus, it is essential to define the phase-space densities as functions
of the coordinates $x^{\mu}=(x^{i},t)$ and the local momenta $\overline{p}^{i}$,
$f(x^{\mu},\overline{p}^{i})$. This choice of variables is particularly
important in keeping the correct scaling properties of the DM density
with the scale factor. In the usual isotropic Robertson Walker background,
the momenta are assumed on-shell, and so the phase-space density depends
on$|\vec{\overline{p}}|$ (i.e. $E$) rather than $\vec{\overline{p}}$.
However, in our case, the small anisotropies that characterize the
foam fluctuations are taken into account by assuming a dependence
on the individual components $\overline{p}_{i}$.

The Liouville operator acting on $f\left(x^{\mu},\,\overline{p}^{i}\right)$
takes the form: \begin{equation}
\hat{L}[f]=p^{\mu}\frac{\partial f}{\partial x^{\mu}}+m\sum_{i}\frac{\partial f}{\partial\overline{p}^{i}}\frac{d\overline{p}^{i}}{d\tau}.\end{equation}
 We apply the isotropy condition , $\frac{\partial f}{\partial x^{i}}=0$,
$i=1,2,3,$ and note that $\frac{d\overline{p}^{i}}{d\tau}=a(t)\frac{dp^{i}}{d\tau}+\dot{a}(t)\frac{dt}{d\tau}p^{i}=a(t)\frac{dp^{i}}{d\tau}+\dot{a}(t)\frac{\overline{p}^{0}}{m}p^{i}$;
the following expression for the Liouville evolution operator is obtained:
\begin{eqnarray}
 &  & \frac{\hat{L}[f]}{p^{0}}=\frac{\partial f}{\partial t}-H\sum_{i}\overline{p}^{i}\frac{\partial f}{\partial\overline{p}^{i}}-2Ha^{2}(t)p^{0}\sum_{i}r_{i}\overline{p}^{i}\frac{\partial f}{\partial\overline{p}^{i}}+8Ha^{4}(t)\sum_{i}r_{i}^{2}\left(\overline{p}^{i}\right)^{3}\frac{\partial f}{\partial\overline{p}^{i}}\nonumber \\
 &  & +\frac{2}{p^{0}}Ha^{2}(t)\sum_{j}\left(\overline{p}^{j}\right)^{2}\sum_{i}r_{i}\overline{p}^{i}\frac{\partial f}{\partial\overline{p}^{i}}+4Ha^{4}(t)\left(p^{0}\right)^{2}\sum_{i}r_{i}^{2}\overline{p}^{i}\frac{\partial f}{\partial\overline{p}^{i}}-4a^{4}(t)H\sum_{j}\left(\overline{p}^{j}\right)^{2}\sum_{i}r_{i}^{2}\overline{p}^{i}\frac{\partial f}{\partial\overline{p}^{i}}\end{eqnarray}
 The mass shell condition gives the energy dispersion relation \begin{equation}
p^{0}=a^{2}(t)\sum_{i}\left(\overline{p}^{i}\right)^{2}r_{i}+\sqrt{\sum_{i}\left(\overline{p}^{i}\right)^{2}+m^{2}}\left[1+\frac{a^{4}(t)\left(\sum_{i}\left(\overline{p}^{i}\right)^{2}r_{i}\right)^{2}}{\sum_{i}\left(\overline{p}^{i}\right)^{2}+m^{2}}\right]^{\frac{1}{2}}\label{energy}\end{equation}
 Further analysis involves the approximation of heavy DM, which we
consider in this work, as the phenomenologically dominant species
are the heavy ones that in general are expected to cause the most
significant distortions in the space time background: $m^{2}\gg\sum_{i}\overline{p}^{i\:2}$.
In this case expanding (\ref{energy}) up to second order and averaging
over the ensembles~%
\footnote{The reason why we can do this at the equation level is because the
time scale of D-particle scatterings is assumed to be much shorter
than the Hubble time $H^{-1}$.%
}, for the random variables $r_{i}$, we obtain for the Boltzmann equation
(now with the binary collision term taken into account): \begin{eqnarray}
 &  & \frac{\partial f}{\partial t}-H\sum_{i}\overline{p}^{i}\frac{\partial f}{\partial\overline{p}^{i}}+6Ha^{4}(t)\sum_{i}\sigma_{i}^{2}\left(\overline{p}^{i}\right)^{3}\frac{\partial f}{\partial\overline{p}^{i}}-\frac{2}{m^{2}}Ha^{4}(t)\sum_{j}\left(\overline{p}^{j}\right)^{2}\sum_{i}\sigma_{i}^{2}\left(\overline{p}^{i}\right)^{3}\frac{\partial f}{\partial\overline{p}^{i}}+4Hm^{2}a^{4}(t)\sum_{i}\sigma_{i}^{2}\overline{p}^{i}\frac{\partial f}{\partial\overline{p}^{i}}\nonumber \\
 &  & -4Ha^{4}(t)\sum_{j}\left(\overline{p}^{j}\right)^{2}\sum_{i}\sigma_{i}^{2}\overline{p}^{i}\frac{\partial f}{\partial\overline{p}^{i}}=\frac{C[f]}{p^{0}}\label{boltz}\end{eqnarray}
To make the connection with \ref{boltzmann} the number density $n\left(t\right)$
is defined as \begin{equation}
n\left(t\right)\equiv\frac{g}{\left(2\pi\right)^{3}}\int d^{3}\overline{p}\, f\left(t,\overline{p}^{i}\right)\end{equation}
 where $d^{3}\overline{p}\equiv d\overline{p}^{1}d\overline{p}^{2}d\overline{p}^{3}$.
Following \cite{Wu} an average temperature $T$ is defined through
the equation \begin{equation}
\frac{g}{\left(2\pi\right)^{3}}\int d^{3}\overline{p}\:\left(\overline{p}^{i}\right)^{2}f\equiv Tmn.\label{temperature}\end{equation}

In the case of superheavy dark matter, the fourth and sixth terms
in (\ref{boltz}) are small compared to the others (and so will be
neglected). Integrating (\ref{boltz}) with respect to $d^{3}\overline{p}$,
yields: \begin{equation}
\frac{dn}{dt}+3Hn=\Gamma(t)n+\frac{g}{(2\pi)^{3}}\int{d^{3}\overline{p}\frac{C[f]}{E}}\label{boltz2}\end{equation}
 where we have incorporated all extra corrective terms in a time-dependent
source term given by: \begin{equation}
\Gamma(t)=Ha^{4}(t)\left(\sum_{i}\sigma_{i}^{2}\right)\left[18Tm+4m^{2}\right]~,\label{source}\end{equation}
and \[
\mathcal{C}\left[n\right]=\frac{g}{(2\pi)^{3}}\int{d^{3}\overline{p}\frac{C[f]}{E}}.\]
The reader should bear in mind that in our analysis we have kept only
leading order terms in the fluctuations. Moreover, the correct scaling
of $n$ with the Hubble parameter, embodied in the term $3Hn$ on
the left-hand-side of (\ref{boltz2}), has been obtained because the
local momentum $\overline{p}^{j}$ has been used in the argument of
the phase-space distribution function $f(x^{i},t,\,\overline{p}^{j})$~\cite{bernstein},
as mentioned earlier.

Hence the rôle of the stochastically fluctuating induced Finsler metrics
in the D-particle foam model is simply to give rise to particle-production
source terms in the Boltzmann equations, linear in the DM density.
The presence of these terms (or equivalently the metric distortions)
will affect quantities such as the thermal relic abundances (which
will be distinguished with a prime). Although formally the situation
is similar to that of time-dependent dilaton and non-critical string
sources encountered in the super critical string (SSC) cosmology model
of \cite{elmn}, our modification $\Omega'_{\chi}$ of the thermal
relic abundance of a single heavy DM species $\chi$ will carry extra
terms as compared to \cite{elmn}. The process we follow is presented
analytically in \cite{msa}. In a standard notation, the result is:
\begin{equation}
\frac{\Omega_{\chi}'h_{0}^{2}}{(\Omega_{\chi}h_{0}^{2})_{{\rm no~source}}}=\left(1+\int_{x_{f}}^{x_{0}}{\frac{\Gamma(x)}{H\, x}dx}-\frac{1}{J}\int_{x_{f}}^{x_{0}}{J(y)\left(\int_{x_{f}}^{y}{\frac{\Gamma(x)}{H\, x}dx}\right)}dy\right)\left\{ \left(\frac{g'_{eff}}{g_{eff}}\right)_{x=x_{f}}\right\} ^{\frac{1}{2}}\label{relic}\end{equation}
 where $J\equiv\int\limits _{x_{0}}^{x_{f}}{\left\langle \upsilon{\tilde{\sigma}}\right\rangle 'dx}$,
$J(x)\equiv\frac{\left\langle \upsilon{\tilde{\sigma}}\right\rangle '}{x^{2}}$
and $x\equiv m_{\chi}/T$, following standard usage; $x_{0}$ denotes
the current value of $x$ corresponding to the CMB temperature $T_{{\rm CMB}}\simeq2.7^{0}K$,
and $x_{f}$ denotes the (D-foam-modified) freeze-out temperature,
estimated by using the freeze-out criterion $Y\left(x_{f}\right)-Y_{{\rm eq}}\left(x_{f}\right)\approx c_{0}Y_{{\rm eq}}\left(x_{f}\right)$,
where $Y(x)=\frac{n(x)}{s}$, with $s$ an entropy density (satisfying
$\frac{d}{dt}\left(sa^{3}\right)=0$) and the suffix {}``$eq$''
denotes the equilibrium expressions (in the presence of the deformed
metric) whose analytic form is given in \cite{msa}. The quantity
$c_{0}$ is a phenomenological constant of order $\mathcal{O}\left(1\right)$,
which can be determined by appropriate numerical fits. An analytic
expression for $x_{f}$ is difficult to obtain in our case \cite{msa};
however, for weak foam effects, approximate expressions can be found
by an appropriate expansion~%
\footnote{For instance, it can be shown that the freeze-out criterion can be
written as~\cite{msa}: $\int_{x_{{\rm in}}}^{x_{f}}{\exp\left(\int_{x_{f}}^{y}{\frac{x'\Gamma(x')}{H_{m}}dx'}\right)\frac{\tilde{\sigma}(y)u_{0}}{H_{m}y^{2}}dy}=(c_{0}+1)^{-1}u_{0}x_{f}^{-3}n_{{\rm eq}}^{-1}(x_{f})$
where $\Gamma(x)$ is the source (\ref{source}), the total cross
section $\tilde{\sigma}(x)\sim\tilde{\sigma}_{0}x^{-j}$, $j=0(1)$
for $s$($p$)-wave annihilators~\cite{kolb}, $u_{0}\equiv\frac{2\pi^{2}}{45}h'~m^{3}$,
where $h'$ denotes the entropy degrees of freedom, and we used $H=H_{m}x^{-2}$,
$H_{m}=1.67g_{eff}'^{1/2}\frac{m^{2}}{M_{P}}$, with $M_{P}$ the
Planck mass; $x_{{\rm in}}$ corresponds to some initial value of
the (inverse) temperature, say at the end of the inflationary era.
The equilibrium number density $n_{{\rm eq}}(x)$ receives corrections
from the foam~\cite{msa}. For weak-foam situations, corresponding
to large values of $M_{s}/g_{s}$ compared to the DM mass $m$, one
may expand the exponential to first order in $\Gamma(x)$, taking
into account that the scale factor $a(t)\sim1/T$ (\emph{c.f}. below),
and solve iteratively for $x_{f}$. On assuming that the fitting constant
$c_{0}$ can be chosen in such a way that the foam corrections on
the right-hand-side of the above freeze-out criterion equation are
subleading compared to their counterparts on the left-hand-side, one
finds self-consistently~\cite{msa} that the freeze-out point $x_{f}$
slightly increases relative to its foam-free-case value, $x_{f}^{(0)}$
(which for neutralino dark matter is of order 20~\cite{spanos}):
$x_{f}^{(j)}\simeq x_{f}^{(0),(j)}+\frac{6}{j+1}\left(g_{s}\frac{m}{M_{s}}\right)^{2}\left(\frac{x_{f}^{(0)}}{x_{0}}\right)^{4}\left(\frac{x_{f}^{(0)}}{x_{{\rm in}}}\right)^{j+1}\left(\sum_{i=1}^{3}\Delta_{i}^{2}\right)$,
where $j=0\,(1)$ for $s$- \, $(p-)$ annihilators, and $\Delta_{i}^{2}$
is a \emph{dimensionless} foam-fluctuation variance (\emph{c.f}. (\ref{defrecvel})).
Hence, the foam leads to a relative decrease of the freeze-out temperature.
However, taking into account that for neutralino masses in the range
of $\mathcal{O}(10^{2})$~GeV~\cite{spanos}, $x_{{\rm in}}\sim10^{-12}$
(since one can reasonably place the end of inflation at temperatures
of order $10^{14}$ GeV), and today's CMB value $x_{0}\sim10^{14}$,
one observes that the shift in the freeze-out temperature is negligible,
for $\Delta_{i}^{2}\le1$, even for low string scales of order TeV.%
}. The quantity $(\Omega_{\chi}h_{0}^{2})_{{\rm no~source}}\equiv\frac{1.066\times10^{9}~{\rm GeV}^{-1}}{M_{P}}\sqrt{g_{eff}}J$
denotes the relic density in the absence of a source term $\Gamma=0$
and $J\equiv\int\limits _{x_{0}}^{x_{f}}{\left\langle \upsilon{\tilde{\sigma}}\right\rangle 'dx}$.
However, the reader should bear in mind that what we denoted as $(\Omega_{\chi}h_{0}^{2})_{{\rm no~source}}$
is not quite the standard expression within the FRW Cosmology, since
$\left\langle \upsilon{\tilde{\sigma}}\right\rangle '$ carries implicit
information about the D-particles effects on the total cross section
(denoted by a prime) and also the freeze-out point that appears as
the upper end of integration is shifted as described above; $M_{P}$
is the Planck mass, which is related to the D-particle mass scale
$M_{s}/g_{s}$ via \begin{equation}
M_{P}^{2}=\frac{M_{s}^{2+\delta}}{g_{s}^{2+\delta}}V_{c}^{(\delta)},\label{extradim}\end{equation}
with $V_{c}^{(\delta)}$ the volume of the compactified $\delta$-extra
dimensions of the 3-brane world, which depends on the specific model
under consideration; the $M_{P}^{2}\left(=1/G_{N}\right)$ enters
the formalism through the effective four-dimensional Einstein-Friedmann
equation $H^{2}=\frac{8\pi G_{N}}{3}\left(\rho+\Delta\rho\right)$.
Here, $\rho$ denotes the total (critical) radiation and matter energy
density, including DM, and $\Delta\rho$ symbolizes collectively all
the stochastic effects of the D-foam on the standard Einstein's equations.
Regarding the latter, we note a small but important difference compared
with the case of the time-dependent dilaton model of \cite{elmn}.
The stochastic fluctuations of the space-time due to the recoil-velocity
fluctuations of the D-particle defects in the D-foam, do contribute
to thermalization of the Universe, indirectly, as a result of their
coupling with the photon or electrically-neutral matter excitations
(such as DM) via the distorted Finsler metric backgrounds~%
\footnote{In our D-foam models~\cite{dfoam,emnnewuncert,li} charged matter
cannot couple dominantly to the D-particles due to electric charge
flux conservation.%
}.

In order to calculate the explicit form of the factor $\left(\frac{g'_{eff}}{g_{eff}}\right)_{x=x_{f}}$
that contributes to the D-foam corrections to dark matter relic abundances,
one should first derive the equilibrium expressions for relativistic
fermions and bosons in the presence of the D-foam. To this end, we
use the basic formula~\cite{msa}: \begin{equation}
\rho=\frac{g}{(2\pi)^{3}}\int{\ll n\omega_{r}\gg d^{3}\bar{p}}\label{dens}\end{equation}
 where $\omega_{r}$ is the equivalent of (\ref{energy}) with $a(t)=1$
and the notation $\ll\ldots\gg$ denotes the average with respect
to the statistical parameters $r_{i}$ of the foam i.e. : \begin{equation}
\ll n\omega_{r}\gg\equiv\prod_{j}\frac{1}{\sigma_{j}\sqrt{2\pi}}\int\limits _{-\infty}^{\infty}{dr_{j}<n\omega>_{r}\exp(-\frac{r_{j}^{2}}{2\sigma_{j}^{2}})}\label{int}\end{equation}
and (for any member of the ensemble of possible foams) we have the
canonical energy distribution\[
<n\omega>_{r}=\frac{\omega_{r}}{\exp(\beta(\omega_{r}-\mu))+\varpi}\]
 where $\varpi=+1\,(-1)$ applies to fermions (bosons) and $\beta=\frac{1}{k_{B}T}$
.

In this way we then obtain the expression~\cite{msa}:\begin{eqnarray}
 &  & \rho=\frac{g}{\pi^{2}}\left(\beta^{-4}\frac{1}{2}f_{3}(1,\beta\mu,\varpi)+\overline{\sigma}^{2}\beta^{-6}\left(\frac{1}{12}f_{5}(1,\beta\mu,\varpi)-\frac{1}{5}f_{6}(1,\beta\mu,\varpi)+\frac{1}{5}\xi f_{6}(2,\beta\mu,\varpi)-\frac{1}{12}f_{6}(1,\beta\mu,\varpi)\right.\right.\nonumber \\
 &  & \left.\left.+\frac{1}{12}\xi f_{6}(2,\beta\mu,\varpi)-\frac{1}{5}f_{7}(1,\beta\mu,\varpi)+\frac{1}{5}\xi f_{7}(2,\beta\mu,\varpi)+\frac{2}{5}f_{7}(1,\beta\mu,\varpi)-\frac{4}{5}\xi f_{7}(2,\beta\mu,\varpi)+\frac{2}{5}\xi^{2}f_{7}(3,\beta\mu,\varpi)\right)\right)\label{equil}\end{eqnarray}
 where $g$ denotes the spin degrees of freedom of either the fermions
or the bosons under consideration and we have set : $\overline{\sigma}^{2}\equiv\sigma_{1}^{2}+\sigma_{2}^{2}+\sigma_{3}^{2}$.
The functions $f_{j}$ appearing in (\ref{equil}) are defined through
integrals of the general form: \begin{equation}
f_{j}\left(l;\mu,\xi\right)\equiv\int_{0}^{\infty}dk\,\frac{k^{j}}{\left(e^{k-\mu}+\xi\right)^{l}}\label{zh}\end{equation}
 that can be related to $\zeta$ and Dirichlet $\eta$ functions.
We also note here that since we are interested in relativistic species
(contributing to $s$) we can set the chemical potential to $0$ ($\mu\ll\omega_{r}$)
in (\ref{equil}) . $g_{eff}'$, the modified effective number of
degrees of freedom, is determined from $\rho_{tot}$, the total relativistic
energy density, since $\rho_{tot}=\frac{\pi^{2}}{30}g_{eff}'T^{4}$;
one finds that: \begin{equation}
g_{eff}'=g_{eff}+\frac{30}{\pi^{2}}\overline{\sigma}^{2}\left(\frac{2\pi^{4}}{189}\sum_{i}{g_{i,b}\left(\frac{T_{i,b}}{T}\right)^{4}T_{i,b}^{2}}+\frac{793.92}{\pi^{2}}\sum_{j}{g_{j,f}\left(\frac{T_{j,f}}{T}\right)^{4}T_{j,f}^{2}}\right)\label{eff1}\end{equation}
 where $g_{eff}$ stands for the standard effective number of degrees
of freedom\cite{kolb} in the absence of D-foam corrections: \begin{equation}
g_{eff}=\sum_{i}{g_{i,b}\left(\frac{T_{i,b}}{T}\right)^{4}}+\frac{7}{8}\sum_{j}{g_{j,f}\left(\frac{T_{j,f}}{T}\right)^{4}}\label{eff2}\end{equation}

The numerical factors appearing in (\ref{eff1}) are the result of
calculating the values of the integrals (\ref{zh}). A detailed analysis
of all these will appear in \cite{msa}. At temperatures of interest
(since a freeze-out temperature is of the order of a few ${\rm GeV}$
for typical dark matter candidates with masses in the range $m\approx1{\rm GeV}-10^{4}{\rm GeV}$
~\cite{spanos}) all the species in the standard model behave as
relativistic matter. A counting of degrees of freedom, then, yields:
$g_{f}=g_{quarks}+g_{leptons}=6\times\left(2\times2\times3\right)+3\times\left(2\times2\right)+3\times2=90$
and $g_{b}=g_{gluons}+g_{EW}+g_{photon}+g_{Higgs}=8\times2+3\times3+2+1=28$.
Substituting these values in equations (\ref{eff1}) and (\ref{eff2})
and using the approximations $\frac{T_{i,b}}{T}\approx\frac{T_{j,f}}{T}\approx1$
we find: $g_{eff}=106.75$ and $g'_{eff}=106.75+22138\overline{\sigma}^{2}T^{2}$.
These considerations imply that the D-foam corrections to the effective
degrees of freedom will be of order $\left(\frac{g_{eff}'}{g_{eff}}\right)_{x_{f}}\approx1+207.38\overline{\sigma}^{2}T_{f}^{2}$.
The dimensionful variances $\sigma_{i}^{2}$ are naturally suppressed
by the (square) of the heavy D-particle mass scale (\emph{c.f}. (\ref{defrecvel}),
with $\ll\xi_{i}^{2}\gg\equiv\Delta_{i}^{2}$, $i=1,2,3$ a \emph{dimensionless}
variance, that could be naturally up to O(1)). Thus, we obtain: \begin{equation}
\left(\frac{g'_{eff}}{g_{eff}}\right)_{x_{f}}\sim1+207.38g_{s}^{2}\frac{m^{2}}{M_{s}^{2}}x_{f}^{-2}\left(\sum_{i=1}^{3}\Delta_{i}^{2}\right).\end{equation}

To determine the final corrections to the relic abundances we may
expand the relevant expressions up to order $\sigma_{i}^{2}$. Expressing
the result in terms of $\Delta_{i}^{2}$, we finally obtain~%
\footnote{We remark at this stage that (\ref{finalrelic}) has been derived
by calculating the integral in (\ref{relic}) under the assumption
that the scale factor $a(t)$ evolves with temperature $T$ as $T\sim\frac{T_{0}}{a(t)}$,
where $T_{0}$ denotes today's temperature. However, this is only
approximately true, since particles' annihilation at different stages
of the evolution of the universe slowed down the cooling due to heat
deposition. Actually, the corrected cooling law, obtained by assuming
that the entropy (which is dominated by the relativistic species contribution)
remains constant despite the presence of the source, reads: $a(t)\approx\left(\frac{g_{eff,0}}{g_{eff}(t)}\right)^{1/3}\frac{a_{0}T_{0}}{T}$.
The function $g_{eff}(t)$ is known ~\cite{kolb} and is such that
$g_{eff,0}\simeq3.36$ (due to the fact that today only photons and
the light neutrinos contribute), while at freeze-out $g_{eff}\left(t_{f}\right)\simeq106.75$
within the framework of the Standard Model. In principle, one should
calculate the integral in (\ref{relic}) numerically using this evolution
of $g_{eff}(t(T))$. For our qualitative purposes in this work we
shall ignore such corrections which, at any rate, do not affect significantly
the order of magnitude of foam contributions to the relic abundances.To
get an estimation of how big the difference is and assuming that the
entropy will be conserved despite the presence of the source term
(see also \cite{elmn}) implies that $g_{eff}a^{3}(t)T^{3}$ will
remain constant throughout the evolution of the universe (only relativistic
species are connsidered to contribute to the entropy). Today only
photons and neutrinos can contribute to the entropy yielding $g_{eff,0}=3.36$.
Therefore very roughly one would get a correction of the order of
$\frac{\left(a(t)T\right)_{GeV}^{3}}{\left(a_{0}T_{0}\right)^{3}}\approx\frac{1}{30}$.
This is fully taken into account in \cite{msa}%
}:

\begin{eqnarray}
\frac{\Omega_{\chi}'h_{0}^{2}}{(\Omega_{\chi}h_{0}^{2})_{{\rm no~source}}}\simeq\left[1+207.38g_{s}^{2}\frac{m^{2}}{M_{s}^{2}}x_{f}^{-2}\left(\sum_{i=1}^{3}\Delta_{i}^{2}\right)\right]^{1/2}\left[1+g_{s}^{2}\frac{m^{2}}{M_{s}^{2}}\,\left(\sum_{i=1}^{3}\Delta_{i}^{2}\right)\left(1+6x_{0}^{-1}\right)\right]\label{finalrelic}\end{eqnarray}
 where clearly the dominant correction terms are of order $g_{s}^{2}\frac{m^{2}}{M_{s}^{2}}\left(\sum_{i=1}^{3}\Delta_{i}^{2}\right)>0$.
As noted in \ref{extradim} it is possible to choose

From the positive signature of the foam-induced corrections we observe
that the relic abundances are \emph{larger} than the corresponding
ones evaluated within a standard cosmological model. In this sense,
the space-time foam background will leave less freedom for supersymmetry
at colliders~\cite{elmn}. This effect is opposite to those induced
by a time-dependent dilaton in supercritical string cosmologies~\cite{elmn}.
In the latter case, the negative signature of the pertinent corrections,
implied that there was \emph{less} DM available today as compared
to the standard cosmology calculations; this could then lead to much
heavier supersymmetric partners produced at colliders, such as the
LHC, with falsifiable signatures~\cite{dutta}. However, the corrections
due to D-foam are in general small, as expected, and they can only
be significant for low string scale models and heavy DM candidates.
Indeed, the string scale $M_{s}$ is a free parameter in the modern
version of string theory (cf (\ref{extradim})). For traditionally
high string scales ($M_{s}\ge O(10^{16})$~GeV), in order for the
D-foam effects to be significant one needs superheavy DM, with masses
higher than $M_{s}/g_{s}$. However, the effects of such a superheavy
DM will be eroded by inflation; moreover superheavy DM would not be
produced significantly during a reheating phase of the Universe after
its exit from the inflationary period. For intermediate string scales~\cite{pioline},
where the quantity $M_{s}/g_{s}$ could be of order $10^{11}$~GeV
(which is the order of the GZK cutoff of ultra-high energy cosmic
rays), there could be significant modifications in the relic abundances
of superheavy DM particles with masses of this order. Such super-heavy
DM particles can be produced during reheating~\cite{riotto}, but
in view of our scenario above, their relic abundance will be modified
from the standard cosmology result. The presence of super-heavy DM,
with increased relic abundances, might provide an explanation for
the production of at least part of the spectrum of the ultra high
energy cosmic rays, with energies of order $10^{20}$~eV. Hence,
the effects of D-foam on such scenarios are worthy of investigating
further, especially in view of the fact that the density of D-particles
might be significantly higher at earlier eras of the Universe, leading
to stronger stochastic effects $O(\Delta_{i}^{2})$.

For low string scales, of order a few TeV, the effects of the D-foam
on thermal relic densities would be more significant. In fact, depending
on the type of DM considered the effect could be constrained or falsified
already by the WMAP five year data, since the induced increase of
thermal relic abundance leaves less room for supersymmetry in the
relevant parameter space. In certain cases, it may exceed the allowed
region set by WMAP. The situation may thus lead to modifications of
supersymmetry searches at colliders, especially in the context of
neutralino DM models with the neutralino being Higgsino- or Wino-like,
with masses up to TeV. In such cases there may be other reasons for
an increased relic abundance, for instance slepton co-annihilation~\cite{slepton},
and in fact our effects of the foam are of comparable strength in
some of these cases.

\bigskip{}
 \textbf{Acknowledgements} The work of N.E.M. and S.S. is supported
in part by the European Union through the Marie Curie Research and
Training Network \emph{UniverseNet} (MRTN-2006-035863) and that of
A.V. by a King's College London Departmental Graduate Studentship.

\end{document}